\documentclass[useAMS,usenatbib,usegraphicx]{mn2e}

\title[The giant star of YY Her ]{The giant star of the symbiotic system YY
Her: Rotation, Tidal wave, Solar-type cycle and Spots}
\author[Liliana Formiggini and Elia M. Leibowitz]{Liliana
Formiggini$^{1}$,$^{2}$ \thanks{E-mail:
lili@wise.tau.ac.il} and Elia M. Leibowitz$^{1}$\thanks{{E-mail:
elia@wise.tau.ac.il}}\\
$^{1}$The Wise Observatory and the School of Physics and Astronomy, Raymond
and Beverly Sackler Faculty of Exact Sciences \\ Tel Aviv University, Tel
Aviv 69978, Israel\\
$^{2}$INAF - Istituto di Radioastronomia,
Via Gobetti 101, 40129 Bologna, Italy}

\def\simi{$\sim $}
\def\apm {$\pm$}

\begin{document}

\pagerange{\pageref{firstpage}--\pageref{lastpage}} \pubyear{2006}

\maketitle

\label{firstpage}

\begin{abstract}
We analyze the historical light curve of the symbiotic star YY Her, from
1890 up to December 2005. A secular declining trend is detected, at a rate 
of $\sim $.01 magn in 1000 d, suggesting that the system could belong to 
the sub-class of symbiotic novae. Several outburst events are superposed 
on this slow decline. Three independent periodicities are identified in 
the light curve. A quasi-periodicity of 4650.7 d is detected for the 
outburst occurrence. We suggest that it is a signature of a solar-type 
magnetic dynamo cycle in the giant component. A period of 593.2 d 
modulates the quiescent light curve and it is identified as the binary 
period of the system. During outburst events the system shows a stable
periodic oscillation of 551.4 d. We suggest that it is the rotation period
of the giant.The secondary minima detected at some epochs of quiescence 
are probably due to dark spots on the surface of the rotating giant.

The difference between the frequencies of these two last periods is the
frequency of a tidal wave in the outer layers of the giant. A period which 
is a beat between the magnetic cycle and the tidal wave period is also 
apparent in the light curve. YY Her is a third symbiotic system exhibiting 
these cycles in their light curve, suggesting that a magnetic dynamo process 
is prevalent in the giant components of symbiotic stars, playing an important 
role in the outburst mechanism of some of these systems.

\end{abstract}
\begin{keywords} binaries: symbiotic -- stars: individual: YY Her -- stars:
magnetic fields -- stars: oscillations.
\end{keywords}
\section{ Introduction}

    The long-term light curve (LC) of a symbiotic star (SS) is quite
irregular, and shows phases of quiescence, with quasi regular brightness 
oscillations  and phases of activity. For a few  SS's, records of their 
photometric behaviour, going back about a century, are available. The 
inspection of these historical LC's reveals a complex photometric 
variability both during quiescence and during outburst epochs. Analysis 
of these  LC's can be used as a tool for gaining insight into the 
symbiotic nature  and may help revealing  properties of the cool
component of these binary systems.

Formiggini \& Leibowitz (1994) analyzed the historical LC of Z And and
discovered  a $\sim $8400 d period for the outburst activity, beside the
758.8 d binary one. During the outburst phases, a period of  $\sim $656 d is 
also present. In BF Cyg, Leibowitz \& Formiggini (2006, hereafter paper I) 
detected, in addition to the well known 757.3 d binary period of the system, 
a 6376 d cycle for the occurrence of the outbursts.
Another significant periodicity of 798 d  was detected for this system, and 
was interpreted as the rotation period of the giant component of this
binary system.  The discovery that outburst events occur with a constant
time interval between them is relevant to the understanding of the origin of 
the outburst phenomenon and of the nature of the clock that regulates their 
appearance. In paper I we suggested that a magnetic dynamo process, similar 
to the well known solar cycle, can be the mechanism that regulates the  
activity events of BF Cyg.

    In this paper we analyze the historical LC of  YY Her, which is among
the prototype symbiotic stars. Its giant component is a M4 star  
(M\H{u}rset \& Schmid 1999). Its optical spectrum shows strong TiO bands, 
but no radial velocity data are available (Kenyon 1986).
The nature of the hot component is not well established. Numerous strong
emission lines are detected in the {\em IUE ( International Ultraviolet 
Explorer)} spectra. The flat UV continuum  can be fitted by a hot main 
sequence accretor (Kenyon \& Webbink 1984). The photometric history of 
YY Her since 1890 has already  been analyzed by Munari et al. (1997),
and a periodic  590 d fluctuation was detected. Several outburst events are
known for YY Her. After the 1997/98 event,  secondary minima appear in 
the B, V, R photometric data, but not in the U light curve 
(Hric, Petrik \& Niarchos 2001). Ellipsoidal variations of the red giant 
have been suggested as the origin of these secondary minima 
(Mikolajewska et al. 2002).

In this work we analyze a 115 years historical LC of YY Her with a
procedure that is similar to the one used for BF Cyg in paper I. 
Although  the photometric data available for YY Her are less regular,
the results are so similar to those found for BF Cyg, that the two stars
may be considered as nearly twin systems.

In Section 2 we describe the data sets used in our analysis and the
strategy adopted for merging the data sets and constructing a time-series 
adequately scaled along the whole time interval of the available 
observations.
In Section 3 we present the time-series analysis techniques used in the
search for  periodicities and the periods detected in that analysis. 
We propose an interpretation of the detected periodicities and of the 
appearance of the secondary minima. In Section 4 we compare the
characteristics of YY Her to those of BF Cyg and discuss some possible
implications on the nature of the giant star in symbiotic systems.

\section { The long-term light curve of YY Her}

The previous study of the photometric history of YY Her by Munari et al.
(1997) is based on two data sets. The first one is of photographic 
magnitudes  from several sources and the second one is a merging of 
visual estimates retrieved from the  AAVSO and AFOEV data banks, 
after correcting for possible systematic errors. Several outburst 
events are evident in this LC but the different scale of the photographic 
and visual samples precludes  a whole overview of the behaviour of 
this system (see Fig. 3 of Munari et al. 1997).

   The photographic magnitudes assembled in Table 6 of Munari et al. (1997)
are sparse data mainly from the Harvard plates collection, and more 
frequent measurements from the Sonnenberg Sky Patrol archive, starting 
from JD 2434099 (1952) up to JD 2439059 (1965). We use these data such 
as published as representative of the ancient LC of YY Her. To these we 
added  the large AAVSO data set of confirmed visual estimates up to 
November 2001 and converted them into means of 10 d.

The last validated data from AAVSO correspond to November 2001. In order to
obtain the LC  up to the present days, we retrieved the data from the AFOEV 
data-bank. We averaged these visual estimates over a time interval of 10 d, 
excluding the upper limits or the uncertain values.
A systematic offset  of .028 magn between this set and the 10-d binned set
from the of AAVSO was estimated comparing the data in the overlapping time 
interval. Scaling the data we were able to continue the  LC for YY Her  
up to December 2005.

\begin{figure}
\includegraphics[width=85mm]{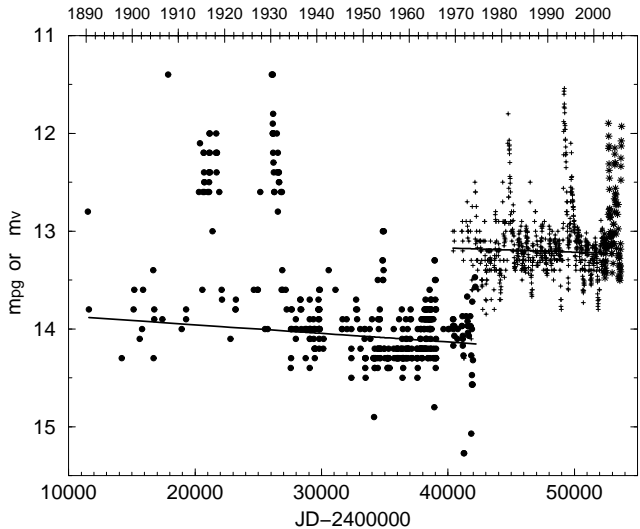}
\caption{A 115 yr LC of  YY Her, from the year
1890 up until Dec 2005. Dots refer to {\em m}$_{pg}$, crosses indicate {\em
m}$_{v}$ data  from AAVSO and stars are scaled {\em m}$_{v}$ from  AFOEV. Solid line
is a linear regression of the quiescence data, in  each of the two data sets.}
\end{figure}

Fig. 1 is a plot of the entire LC obtained from all our data sources. Two
distinct subsets are clearly seen in the figure.
In order to establish the {\em m}$_{pg}$ and the {\em m}$_{v}$ LC's on a
common zero-point level we considered the few measurements that populated 
the overlapping portions of the two light curves. Comparing the magnitudes of 
points that are close in their time of observations in the two LC's we derive 
the  value ({\em m}$_{pg}$-{\em m}$_{v}$)= .90.

A more accurate value has been obtained by also taking into account  the
general trend of the system during quiescence state, in each of the two 
light curves separately. For this purpose, we eliminated  from both samples 
the measurements classified as uncertain and all the brightest data belonging 
to epochs of activity. Applying a linear regression to the quiescence state 
sections of each of the two light curves, a slow decline with time is 
detected in both of them. The slope of the two lines in Fig. 1 is  
.96 10$^{-05}$ and 4.3 10$^{- 06}$ for the {\em m}$_{pg}$ and the {\em m}$_{v}$ 
data respectively.  The fact that a negative slope is measured in the two
independent LC's, at two different time intervals, is evidence of the
reality of the trend. Considering the statistical uncertainty in the slope, 
due also to the different time-length of the two data samples, these result 
are consistent with a general decline at a constant rate throughout the entire 
time of the observations. The no trend result reported by Munari et al.
(1997) is probably due to the relatively short time interval (1969-1993)
considered by these authors. Our second subgroup of data points that does 
show the decline includes twelve additional years, based on AAVSO data.

When the decline trend is taken into account, the scaling factor between
the two lines in Fig. 1 in the overlapping regions of the two data sets
takes the value  .97 magn, and we applied it to the AAVSO data. We  then
calculated a linear regression to the complete sample of quiescent data. 
The resulting rate of the steady decline of the system from 1890 up to 
November 2001 is  $\sim $ .01 magn in 1000 d. The trend of decline 
detected here, suggests that the system is recovering from a major
brightening event, which occurred sometime before  1890, the year of the
first recorded magnitudes. A decreasing trend in the long-term LC is one 
of the characteristic of the sub-class of symbiotic novae such as 
HM Sge, RR Tel or  BF Cyg. This behaviour is associated with a single 
major nova-like event that occurred  in these systems, followed by a 
very slow fading that lasts for more than a century.

The dots in Fig. 2 show the measurements of YY Her from 1890 up to Dec
2005, de-trended for the secular decline as explained above. The total 
time covered by the observations is \simi 115.6 yr, but the distribution 
of data within this time range is not homogeneous. Due to the scarcity of 
points of the very old observations, in the time series analysis that we 
applied to the data we have disregarded all points prior to JD 2420000. 
Therefore the analysis is applied to a 92.37 years LC. 
Several outbursts, with different amplitudes, are clearly seen in this LC, 
although only the last two  are well covered by frequent measurements. 
We shall address the bias due to the non homogeneous distribution of 
points in the LC that we have analyzed.

\begin{figure}
\includegraphics[width=80mm]{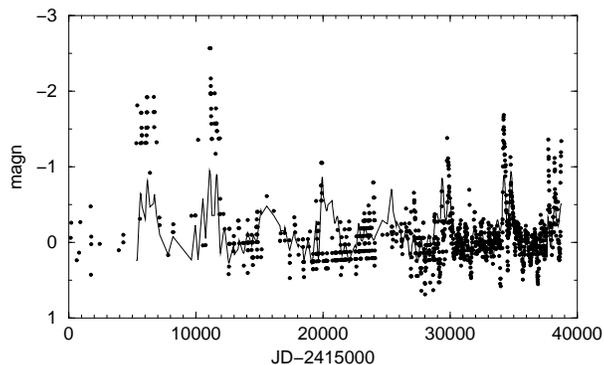}
\caption{The de-trended LC of  YY Her. The line is the artificial LC 
calculated as explained in Section 3.5.}
\end{figure}

\section{Time Series Analysis}

\begin{figure}
\includegraphics[width=90mm]{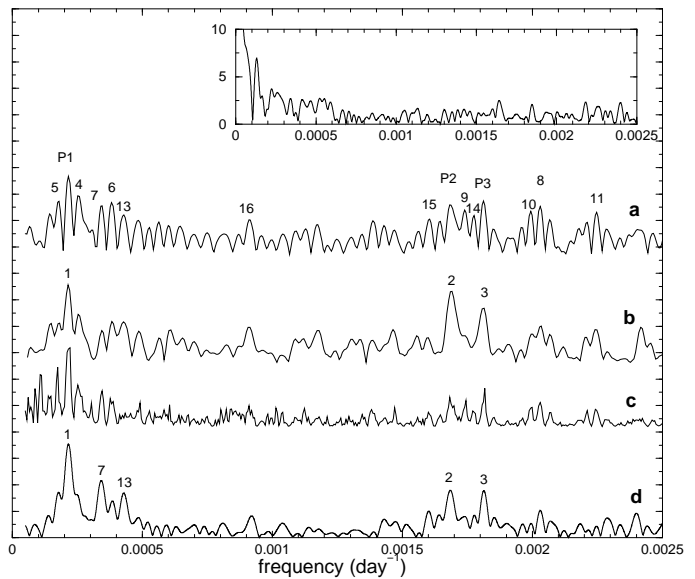}

\caption{The power spectrum  (PS) of the LC of  YY Her shown in Fig. 2.
The insert is the window function.
(a) The  PS of the observed LC according to the Scargle algorithm, 
(b) The  CLEAN PS, (c) The PS  according to the  AoV method, (d)
The periodogram of the artificial LC calculated as explained in Section 3.5.}
\end{figure}

\subsection{General Periodic Content}

In order to examine the overall  periodic content of the LC we begin our
analysis by computing the power spectrum (PS) of the LC shown in Fig. 2  
by means of the Lomb-Scargle algorithm (Scargle 1982). The PS is 
presented  in Fig. 3 (a); the insert shows the window function.
Due to the very uneven distribution of the observation points along the 
time axis, many of the peaks in the PS do not represent genuine 
periodicities in the LC but are rather aliases of other periods. In order 
to clean the PS from these spurious peaks we applied the CLEAN algorithm 
for time-series (Roberts, Leh\'{a}r \& Dreher 1987). Fig. 3 (b) is the 
CLEAN PS. The PS in both Fig. 3 (a) and (b) is dominated by a highly 
significant peak designated {\em P}1 in the figure. The 2$^{nd}$ and 
3$^{d}$ highest peaks are marked as {\em P}2 and {\em P}3. The other 
peaks in the  PS are also marked by numbers ordered by their height.

From inspecting Fig. 2 it is quite clear that the cycle of {\em P}1 in the
LC is highly non-harmonic. Since the Lomb-Scargle PS search technique 
(Scargle 1982) is especially sensitive to harmonic signals, we applied 
on the LC the Analysis of Variance (AoV) (Schwarzenberg-Czerny 1989) 
method. This technique does not favor any particular structure in the 
search for cycles. Fig. 3 (c) is the AoV periodogram, where the three 
dominating features are the same highest peaks in the PS seen in frames 
(a) and (b). Thus it is clear that the three periodicities have not 
been introduced into the data by the algorithm of the analysis.

\subsection{The  {\em P}1 periodicity}

The period corresponding to the  {\em P}1 peak is \simi 4638 d. We applied 
the bootstrap statistical test (Efron \& Tibshirani 1993) on the  LC. 
It showed that the probability of obtaining in the PS a peak as high as 
the {\em P}1 from a random distribution of magnitudes at the times of the 
observations is smaller than 10$^{-3}$.

In order to strengthen even further the reliability of the statistical
significance of the {\em P}1 peak we repeated the bootstrap test with 
the following procedure.
We bin the LC onto 128 bins, each of 300 d width. With this binning, each
time interval corresponding to the suspected binary period of the system
(\simi 600 d, Munari et al. 1997) is represented by just two points. The
outburst events themselves are represented by just three to seven points. 
We compute the PS of the resulting time series. The highest peak in this 
LC corresponds again to the {\em P}1 period. The bootstrap test indicates 
that even for such a poorly sampled LC, the probability to obtain in the 
PS, as a random event, a peak as high as that of the {\em P}1 periodicity, 
is less than 1/300.

The periodic or quasi-periodic nature of the series of outbursts of YY Her
therefore seems to be statistically highly significant.

Fig. 2 shows that the cycle of the {\em P}1 periodicity  does not have a
stable structure. In particular the amplitudes of the  outbursts are very 
different  from one cycle to another. The value of {\em P}1 therefore 
should not be construed as the period of a coherent oscillation, but 
rather as the mean value of a rather regular time interval that separates 
successive outbursts of the system.
The meaning of {\em P}1 is therefore similar to the meaning of the
11.3 year interval that is commonly referred to as the period of the Solar
activity cycle. As is well known, the actual time interval between
successive minima of the solar cycle varies between 9.5 and 12.5 years
(Lorente \& Montesinos 2006), or even between 7 and 17 years (Rogers,
Richards \& Richards, 2006).  Further discussion of the nature of the 
{\em P}1 cycle and its relation to the {\em P}3 periodicity is presented 
in Section 3.7.

Fig. 4 (a) displays the LC of the star, superpose on a Sine wave of the {\em P}1
periodicity. The figure shows that while the structure of individual
outbursts varies considerably, the outburst repetition time {\em P}1 does
indeed represent a strong periodic modulation of the timing of the 
phenomenon.

\begin{figure}
\includegraphics[width=80mm]{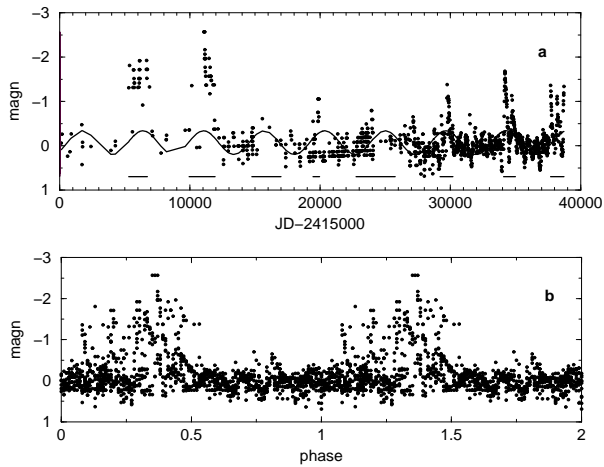}
\caption{ (a) The thick line is a sin wave of periodicity {\em P}1, 
superposed on the LC of YY Her. The segments shows the time intervals visually 
selected as belonging to the active states (see section 3.3).
(b) The detrended LC of YY Her folded onto the period {\em P}1=4650.7 d.}
\end{figure}

Fig. 4 (b) displays the LC of the star folded onto the {\em P}1=4650.7 d period.
This particular value for the period {\em P}1 is explained in Section 4.3.
The complete absence of high points in about one half of the {\em P}1 cycle is
again a clear demonstration of the cyclic nature of the outbursts phenomenon in
the YY Her system.

\subsection{The {\em P}2 and {\em P}3 periods}

The two highest peaks in the high frequency part of the PS, numbered 2 and
3 in Fig. 3 (a), and particularly distinguishable in Fig. 3 (b), correspond to 
the periods  {\em P}2\simi594 d and {\em P}3\simi551 d. The {\em P}2 
period was already identified  in the LC of YY Her and it is commonly 
considered the binary orbital period of the system (Munari et al. 1997). 
We note, however, that so far we are unaware of spectroscopic measurements 
that confirm this identification. The peak of the {\em P}3\simi551 d 
periodicity is nearly as prominent in the PS as that of the {\em P}2 period.
In order to show even better their prominence and independence we proceed as follows.
We compute a running mean LC (RMLC) by applying the running mean operator 
on the observed LC. The width of the running window
that we used in this operation is 600 d, close to the 594 d suspected
binary period of the system, as mentioned above.
In the low frequency range, up to  f=0.001 (1/day) the PS of the RMLC is
identical to that of the observed LC. It shows that the low frequency
brightness variations of the star are independent of the high frequency
variations. We then consider the residuals of the observed LC (RSLC) obtained by
subtracting the RMLC from the observed one. This is the LC of the star from
which the low frequency variability, in particular the {\em P}1
periodicity, is removed. Note that by subtracting the RMLC we remove the 
low frequencies without using, or even assuming, any cyclic behavior of 
the star on the $\ge$ 600 d time scale.

Fig. 5 (a) is the CLEAN PS of RSLC. The {\em P}1 periodicity is absent and
the two domineering peaks of {\em P}2 and {\em P}3 are highly 
significant. However, these two periodicities are not contemporaneous in
the LC of the star. 

\begin{figure}
\includegraphics[width=80mm]{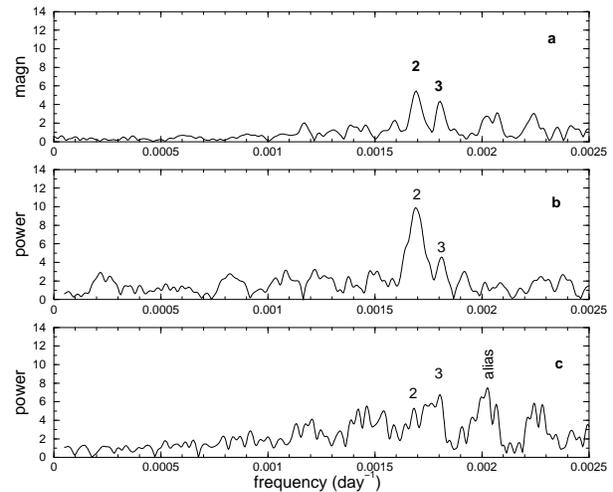}
\caption{ (a) The CLEAN PS of the residual running mean LC (RSLC) of YY Her. 
(b)  The PS of the low-state points L$_j$ RSLC,  (c) The PS of the
high-state points H$_j$ RSLC.}
\end{figure}

In order to show this, we consider
separately two subsets of the LC points. One is the High (H)LC,
consisting of points measured during outbursts of the star. The Low (L)LC
consists of the observed points during quiescence states of the system.
The division between H and L LCs was done in three different ways: (1)
Applying a brightness criterion in the selection of the points. 
In particular, an H$_m$ LC is constructed from all points with magnitude 
brighter than -0.5 mag in Fig.2. The complementary low points constitute 
the L$_m$ LC. Our results are similar when we consider any division threshold 
between the H$_m$ and the L$_m$ LCs in the range -.2 to -.6 mag. 
(2)  Determining by eye in Fig. 2 the time intervals at which an outburst of 
the star is apparent. These time intervals are indicated by the heavy line 
segments along the x-axis of Fig. 4a. All points of the RSLC that fall 
within these intervals are the H$_j$ LC (j for JD determined). The 
complementary points of the RSLC constitute the L$_j$ LC. 
(3) Selecting points by phase in the {\em P}1 cycle.
This is done with the help of Fig. 4b. The H$_p$ LC consists of all RSLC points
that fall between phase 0.1235 and 0.5635 in the {\em P}1 cycle, where 
phase 0 is taken arbitrarily to be on JD 2415000. All points in the 
complementary phase interval are the L$_p$ LC.

The PS of the L$_m$ LC (not shown in this paper) has a distinct, highly
significant peak corresponding to the {\em P}2 periodicity with hardly a
trace of the {\em P}3 period in it. The PS of the H$_m$ LC is very noisy, 
due to the scarcity of points in that subset of the data. We therefore 
consider now the L$_j$ and H$_j$ LCs. Fig. 5 (b) is the PS of the 
L$_j$ LC. It is very similar to the PS of the L$_m$ LC. Fig. 5 (b)
shows the dominating, highly significant peak of the {\em P}2
periodicity. Fig.5 (c) is the PS of the H$_j$ LC. Here the second highest
peak is {\em P}3 (the highest one corresponds to {\em P}=492 d which is 
a 4650 d alias of P3, introduced into the data by the selection process
of the H$_j$ LC). The {\em P}2 peak in this figure is hardly above 
the noise level of the curve. This shows that during outbursts, the 
star oscillates mainly with the {\em P}3 periodicity, while the 
{\em P}2 period plays only a secondary role.

The {\em P}3 peak emerges from the noise in this PS while as mentioned
above it is not prominent in the PS of the H$_m$ LC. The reason is that 
in the selection criterion of the H$_j$ LC, we include also faint points 
that are measured during outbursts, while we select them out from the 
H$_m$ LC. Thus the H$_j$ LC includes complete oscillation cycles of the 
star during outbursts, and therefore the PS is able to discover them. 
The H$_m$ LC, on the other hand, includes only the brightest points 
that constitute only part of the oscillation cycles. Therefore, the P3 
periodicity is not well identified by the PS of that LC.

Finally, we note that the power spectra of the L$_p$ and H$_p$ LCs, namely,
the Low and High LCs obtained by selecting points according to their 
phase in the {\em P}1 cycle, are virtually the same as those shown in 
Fig. 5 (b) and (c).

\begin{figure}
\includegraphics[width=90mm]{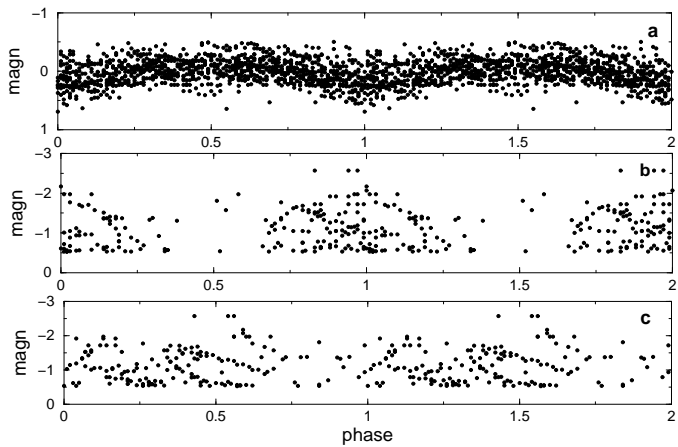}
\caption{ (a) The folding of L$_m$ LC onto {\em P}2 =593.2 d.
(b) The folding of H$_m$ LC onto {\em P}3 =551.4 d. (c) The folding 
of H$_m$ LC onto {\em P}2 =593.2 d.}
\end{figure}

Figure 6 (a) is the folding of the observed  L$_m$ LC onto the period 
{\em P}2. The systematic variation of the luminosity of the system in 
its quiescent  state with this periodicity is seen unmistakably. Note 
that the period, as well as the phasing, are stable throughout the 
entire 92.37 years of the observations. They are undisturbed by the 
violent events of the outbursts that interrupted seven times the 
relative calmness of the system.

Fig. 6 (b) displays the observed H$_m$ LC, folded onto the {\em P}3 period. 
It shows clearly that all the time points at which the star was brighter 
than -0.5 mag are concentrated within one and the same half of the 
{\em P}3 cycle. This demonstrates rather clearly that the brightening of 
the star in its outbursts occurs periodically also with a stable period 
{\em P}3 throughout the \simi92.37 years of the observations.

For comparison we show in Fig. 6 (c) the H$_m$ LC folded onto the {\em P}2
periodicity. Here the distribution of the points is nearly homogeneous,
indicating that the oscillations during outbursts are not modulated by the
{\em P}2 period, the one that modulates the low state of the system.

\subsection{Combination periods}
Among the high peaks in the low frequencies end of the CLEAN PS of YY Her
we found that peak No.13 corresponds to the period {\em P}13 \simi 2300 d
which is the second harmonic of {\em P}1. We also found that {\em P}7 is
the beat period of {\em P}1 and the beat period of {\em P}2 and {\em P}3.
In frequency unit the relation is:  f7 = f1-(f2-f3). We shall return to 
this relation in the following sections.

\subsection{Period value determination}

As described in Section 3.1, the low frequency oscillations of the star are
dominated by the {\em P}1 periodicity, with an amplitude and structure that
vary in time. The high frequency oscillations are dominated by the {\em P}2
and {\em P}3 periodicities, both having stable frequency throughout the
entire 92.37 years time interval covered by the observations. While the
frequencies are stable, the amplitudes are clearly not. Their ratio at low
state are very different from the ratio in the high state.

In order to better determine the value of the three independent periods
that we identify in the LC, we proceed as follows.

We consider three synthetic LC's. One is a two term Fourier series with
the {\em P}2 and {\em P}3 periodicities, representing the H$_p$ LC. The
second one is another two term Fourier series with the same pair of
periods, representing the L$_p$ LC. The third one is a three term Fourier
series constructed with the periods {\em P}1, {\em P}1/2 and the period
1/(f1-(f2-f3)), as explained above, representing the RMLC. We look for 
the best simultaneous fit of the sum of
these three synthetic LC's to the observed one, in the least squares sense, 
from JD 2420000.
In this fitting procedure there are three independent parameters, {\em P}1,
{\em P}2 and {\em P}3.

Best fit is obtained with the following three values: {\em P}1=4650.7 \apm
35 d, {\em P}2=593.2 \apm 1.2 d and {\em P}3=551.4 \apm 1.4 d, that also
imply  {\em P}13=2325.4 d and {\em P}7=2917 d. The uncertainty estimates
are explained in Section 3.6.

Similar results are obtained when the subdivision of the residual LC
between HIGH and LOW curves is done by JD selection rather than by the 
{\em P}1 phasing (see Section 3.3).

The solid curve in Fig. 2 is the synthetic LC that is constructed with the
above values of the three independent periods {\em P}1, {\em P}2 and {\em
P}3 and of the two combinations made up from them, {\em P}7 and {\em P}13.
Fig. 7 (a), (b), and (c) are blowups of three sections of the LC of Fig. 1,
enabling a better appreciation by eye of the quality of the fit.

\begin{figure*}
\includegraphics[width=97mm]{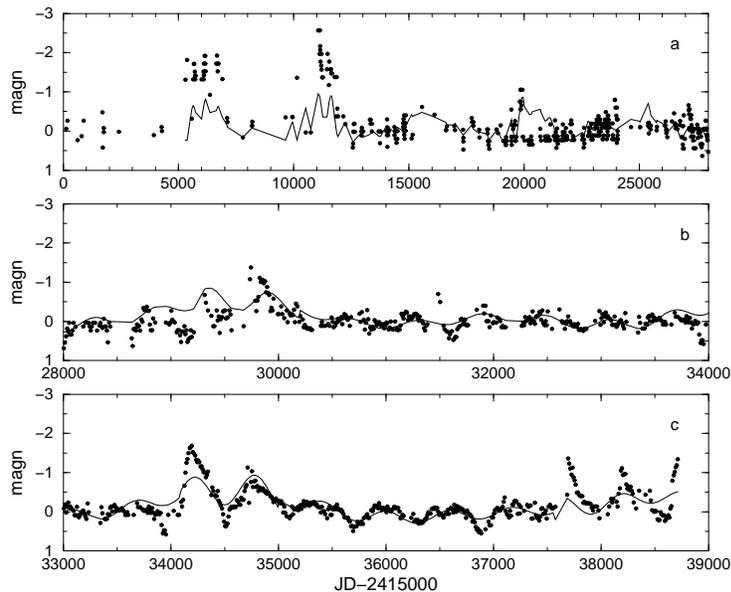}
\caption{Dots are the observed light curve.  Solid line is the synthetic
light curve constructed with the periods {\em P}1, {\em P}2 and {\em P}3
and the two combination {\em P}7 and {\em P}13. (a), (b) and (c) are 
zooms on different  sections of the LC.}
\end{figure*}

The curve in Fig. 3 (d)  is the {\em P}S of the synthetic LC depicted as
a solid line in Fig. 2 and 7. The discrete points used in the computation
of this PS are the calculated magnitudes at the sampled times in the
observed LC. There is a great similarity  between this curve and the 
CLEAN LC of the observed data, curve b in Fig. 3.
In particular, we found that 16 out of the 20 highest peaks in the PS of the 
observed LC have 16 counterparts among the 20 highest peaks in the PS of 
the synthetic LC. In  counterparts, we  mean frequencies that differ from each
other by less than the inherent uncertainty in the frequency  values which is 
 1/T, T being the length (in days) of the 92.37 years of the LC.

The conclusion of this section is that it is possible to construct with
merely three independent periods, {\em P}1, {\em P}2 and {\em P}3,  a time
series that is very similar in its temporal behavior to the observed LC of YY
Her.  These results permit us to conclude that the presence of the three 
periods {\em P}1, {\em P}2 and {\em P}3, as well as {\em P}7 and {\em P}13 
in the LC of YY Her may be considered as established (and see also 
Section 3.7).

\subsection{Accuracy  of {\em P}1, {\em P}2 and {\em P}3}

In order to estimate the accuracy in our derived values of the three
periods {\em P}1 {\em P}2 and {\em P}3 we performed a bootstrap analysis on
the
data (Efron \& Tibshirani 1993). We subtract the value of the theoretical
LC
from the observed one at each time point
of the observations. The set of the differences so created may be
regarded as a sample of the noise values in the observations. We now
construct a pseudo-observed LC by adding to each of the theoretical LC
one number chosen randomly from this sample. On this pseudo-observed LC
we apply the same period determination procedure that we have applied on
the observed one, as described above. We obtain a set of three values for
the three period {\em P}1, {\em P}2 and {\em P}3. We perform this process
repeatedly  1000 times and obtain 1000 triplets of {\em P}1, {\em P}2 and
{\em P}3 values. The central interval of a histogram that includes 950 of
the
{\em P}1 values so obtained may be considered the 95 percent uncertainty
interval
around the {\em P}1 value derived from the observed LC.
The same goes for {\em P}2 and {\em P}3. These results are the period errors 
quoted in Section 3.5.

\subsection{Stability of {\em P}1, {\em P}2 and {\em P}3}

In view of the results presented in Section 3.2, particularly the
folding of the observed LC as shown in Fig. 6 (c) and 6 (d), we may
conclude that the two periods {\em P}2 and {\em P}3 preserve their 
frequency throughout the 92.37 years of the monitoring of the star. 
Each one of these cycles all but disappeared from the LC at certain 
epochs along the history, always reappearing at the same phase. 
Thus, the {\em P}3 period that dominates the oscillations during 
outbursts seems to be squelched at quiescence epochs. When it returns 
to the LC at the next outburst it appears with the same phasing as before. 
The same goes for {\em P}2. Upon returning to the LC after an outburst 
it appears with the same phase as before the outburst.

In order to check even further this claim we divided the entire LC into two
distinct subsets, one of all points up to  JD2440000, and the other of all the
later points. The PS of each of these subsets is of course noisier than
those presented in Fig. 3, especially that of the older set of the less
numerous observations. However, the two peaks of the P2 and P3 periods are
clearly distinguishable in both plots. Also the folding of the two subsets
on either the P2 or the P3 periodicity show the same phasing as in Fig. 6.

These results indicate that each one of the clocks in the system that is 
responsible for one of these two cycles keeps a stable frequency, 
notwithstanding the dramatic variations that the system is undergoing, 
e.g. in its luminosity. We therefore suggest that the two clocks are 
the binary orbital cycle, as suggested by others (Munari et al. 1997), 
and the rotation period of the giant star of this system.

The {\em P}1 periodicity does not share the stability that characterizes the
{\em P}2 and {\em P}3 periods. In fact an outburst event of the system seems
to consist not of a rise in the DC output of the system but rather of a
series of one to three violent oscillations of amplitudes that are significantly
larger than the amplitude of the binary cycle. The fact that an outburst of
YY Her is actually a series of violent oscillations is evident particularly
in the later events, where the observational coverage is more dense (see
Fig. 2). In these cases we see in particular that during outbursts, the
brightness of the system is sometimes falling even below the mean brightness
level of the quiescence state.  Fig. 6 (b) shows that these oscillations have
the {\em P}3 (and not {\em P}2) periodicity. The non stability of {\em P}1 is 
at least partly due to the fact that the {\em P}3 is indeed stable. Since the 
two periods are not commensurate, the onset of the outburst is not stable in 
the {\em P}1 phase space. Furthermore, as already mentioned, the number of 
{\em P}3 oscillations within a given outburst event, varies between one and 
three. This is another cause of the non stability of the {\em P}1 cycle. 
  
The amplitude of the {\em P}3 oscillations within an outburst event is not
constant and hence also the overall amplitude of the outburst events is
clearly varying among the different cycles. Our reconstructed LC is unable
to mimic the large upsurge in the luminosity observed in the first two
recorded outbursts. We do note, however, that at least part of this
discrepancy may be due to unknown errors in the measurements, as well in the
zero term in the scaling of these observations that were made so many
decades ago.

As discussed in Section 3.2, the {\em P}1
periodicity should be regarded as an average value of the quasi-periodic
repetition time of the outbursts of the YY Her system. The meaning of the
{\em P}1 period in the context of YY Her is therefore similar to the
meaning of the well known and much in use period of 11.3 year of the
solar activity cycle (see Section 4).

\subsection {The Secondary Minima}

The LC of YY Her contains additional evidence for the P3 periodicity, as
well as for its interpretation as the rotational period of the giant
component. In a series of photoelectric measurements performed along the
last few binary cycles of the system, Hric et al. (2001) identified in two
cycles secondary minima that appear in between successive primary minima.
Mikolajewska et al. (2002) interpreted this signal as a trace of the well
known ellipsoidal effect in binary stars.

In order to investigate this interpretation we consider now in details the
last four cycles of the binary revolution before the onset of present day
outburst, that are well covered observationally. We combine the photoelectric 
measurements of Munari et al. (1997), Tatarnikova et al. (2000), Hric et 
al. (2001), Mikolajewska et al. (2002) with the AAVSO published confirmed 
magnitudes.  Fig. 8 presents the running mean of this LC with a 60 d wide 
running window. The arrows designate times of successive primary and secondary 
minima. The asterisks and the circles on the x axis of the figure indicate 
the times of the primary and of the secondary minima, respectively.

In the ellipsoidal scenario, the secondary minima in the LC of a binary
system are seen when the longer axis of the elliptically structured giant
star is pointing in the direction of the observer. At this phase the
surface area of the giant that emits in this direction takes a minimum
value. The bulge in the outer layers of the giant has a fixed direction in
the binary rotating frame. Therefore, the frequency of the ellipsoidal
effect should be the frequency of the orbital revolution, or rather twice
this value.

The light variations due to the ellipsoidal effect are expected to be
gradual entrance to and exit from the two minimum phases in the binary
cycle. The structure of the LC seen in Fig. 8, in particularly around the
secondary minima, is hardly having this structure.

\begin{figure}
\includegraphics[width=80mm]{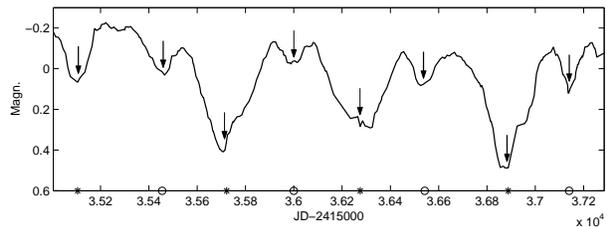}
\caption{ (a) The four last cycles  of YY Her. The line is a 60 d running
mean curve
of the AAVSO and photoelectric data. Stars and circles on the x axis mark
the times
of the primary and of the secondary minima, respectively.}
\end{figure}

The arrows indicating the minima in Fig. 8 were positioned at mid points
of the corresponding minimum profile, as judged by the eye. When we repeat
the eye positioning a number of times, determining again the mid points
independently, we find that the variations in the minimum times so
determined are no larger than $\pm$6 d. The uncertainty in the computed
mean time between successive minima is of the same order.

We find that the mean time difference between successive primary minima is
594.3 d, practically the same value of {\em P}2 derived for the quiescence
state from the entire LC. The mean difference between secondary minima is
561.7 d, consistent with {\em P}3, and significantly different from {\em
P}2. When we take as the time of each minimum simply the day of the lowest
brightness value of the corresponding minimum event, the mean differences
take the values 588 d and 557 d, respectively. This result seems to prove
that the ellipsoidal effect cannot be responsible for the secondary minima.

Consistently with our previous suggestion that {\em P}3 is the rotation
period of the giant, we propose here that the secondary minima result 
from star spots on the surface of the giant. For an a-symmetric lateral 
distribution of spots, as is the case for the sun and  other stars, e.g.  
spotted G and K giants (Bopp \& Rucinski 1981), periodic light variations
at the rotation frequency are indeed expected. As stellar spots are not 
permanent features on the surface of the star, their light modulation at 
the rotation frequency, in amplitude as well as in phase, is expected to 
vary between different activity cycles of the star, and to some extent 
also within one and the same active cycle. This {\em P}3= 551.4 d 
modulation may also disappear completely from the LC, as was the case, 
up to the accuracy of the observations, during the previous quiescence 
state of the star. This would happen when the spots are distributed
more evenly along the latitude coordinates of the rotating star.

Finally, we note that evidence for the {\em P}3 periodicity revealed in the
time of the secondary minima is independent of the evidence for this 
periodicity that we found with our time series analysis presented in 
Section 3.3. There the P3 periodicity manifests itself mainly during 
the outburst state of the system, and through its beats with the
other two periods of the system, as explained in that section.
Indeed, the theoretical LC presented as the solid line in Fig. 2 and 7 
does not show the secondary minima of the last quiescence time interval 
of the star. These minima appear in the LC due to the effect of the 
dark spots, which is not taken into account in our analysis of the LC 
in its entirety. This is quite appropriate since the spots are not 
always there and even when they appear, they do not preserve
the phases of maxima and minima in the LC in different active cycles. 
The analysis presented in this section therefore provides additional, 
independent evidence for the {\em P}3 periodicity.

\section {Discussion}

The analysis performed in  Section 3, revealed the presence of
three independent  periodicities in the historical LC
of YY Her. The period {\em P}1\simi 4650.7 d regulates the  activity
phases of YY Her. During the quiescence state of the system, the
variation is modulated mostly with the {\em P}2\simi 593.2 d period,
which is interpreted
as the binary orbital period. However, the {\em P}3\simi 551.4 d period
is also apparent in the last epoch of quiescence of the star through the
secondary minima superposed on the {\em P}2 modulation. During the outburst
epochs the system enters an oscillation mode with the {\em P}3 period, with
an amplitude reaching 2 magn or more. This amplitude is more than twice the
amplitude of the {\em P}2 orbital modulation.
 We interpret this modulation as due to the rotation period of
the giant component, similar to the results for BF Cyg (paper I).

The thousand  days time-scale of the {\em P}1 period is  similar
to that  already detected for two symbiotic systems, namely for Z And,
\simi 8400 d (Formiggini \& Leibowitz 1994) and more recently 
\simi 6376 d for BF Cyg (paper I) . We already proposed the existence
of a magnetic cycle in the cool giant component as the mechanism
that regulates these multiple outbursts. In analogy with the sun and
other late type stars, a magnetic dynamo process suggests itself as
the cyclical generator of surface magnetic fields that regulate the
activity of the star and in particular the mass-loss rate and the
dynamics of the star wind. Note that  magnetic activity in the
giant component of symbiotics has already been invoked by Soker (2002)
from a theoretical point of view.

There is theoretical and observational evidence for the presence of
magnetic fields in some late-type giants, such as asymptotic-giant-branch
stars (AGB). Observationally, magnetic fields are detected as the 
source of maser polarization found around  AGB stars and of the X-ray 
emission measured for cool giant stars (Ayers, Brown \& Harper 2003).
Invoking dynamo action is also supported by theoretical models  (Soker
\& Kastner 2003, Dorch 2004). Among many modes of dynamo action,
Dorch (2004) recognizes an  exponentially amplification of the magnetic
field on a time scale of about 25 years.

This is  the time scale of the outburst periodicity found by us for the
three SS's Z And, BF Cyg and YY Her.

\subsection{Detailed Comparison with BF Cyg}

In paper I we presented an analysis, similar to the analysis presented
in this work, of a 104 year historical LC of the symbiotic star BF Cyg.
As already mentioned in Section 1, BF Cyg and YY Her are quite similar
to each other in some of their characteristics. Table 1 lists some of
these features and gives their value in the two systems. Here we draw
attention in particular to the similarities in the kinematical
parameters.

The giant component in the two systems is rotating nearly synchronously
with the orbital revolution. For BF Cyg, this result is consistent with
the conclusion of Zamanov et al. (2006) on the rotation of the giant stars
in symbiotic systems. The star YY Her was not investigated in that study.

Although from the point of view of the temporal evolution of
circularization and synchronization of stellar binary systems, BF Cyg
and YY Her may be considered as reaching synchronization, our results
for the two stars indicate that in both cases a complete locking of the
star spin rate has not been achieved yet. The difference between the
spin and the orbital frequencies is less than 10 percent of their value. 
It is nonetheless rather significant, and has a profound effect on the 
physical processes that take place in the outer envelope of the giant.

The two systems differ from each other in the direction of the deviation
from a full synchronization. In BF Cyg, the spin period, 798 d, is longer
than the orbital period, 757 d. In YY Her the spin is faster. Its
period is 551 d, as compared to the 593 d of the orbital period.

The gravitational pull of the hot component in the two stars excites a
tidal wave in the atmosphere of the giant. If we assume that the rotation
of the giant and the orbital revolution are in the same direction, then
in the rotating frame of the BF Cyg giant, the sense of the tidal wave
is pro-grade. For an observer on the surface of the giant of YY Her, on
the other hand, the tidal wave is traveling in the opposite direction
to the spin and to the orbital motion. This is indicated by the negative 
sign in Table 1.

A tidal wave in the outer layers of the giant is an additional,
periodic, mostly radial motion in the equatorial plane of the atmosphere 
of the star. This is another important flow field, in addition to 
the differential rotation, the convection and the meridional 
flows that are considered responsible for the 11/22 years 
magnetic dynamo process in the sun (for a recent
comprehensive review of models of the solar magnetic dynamo processes
see Ossendrijver 2003). The sun itself does not possess it.
In the two symbiotics, the effect of this
additional motion on the dynamo process is manifested by additional
modulation on the activity cycle of the stars, at the beat period of the
cycle fundamental frequency f1 and the frequency of the tidal wave in
the giant rotating coordinate system. Note that the beat frequency in
both cases is the algebraic sum of the cycle frequency and the tidal
wave frequency. In BF Cyg, where the tidal wave has a prograde motion,
the two frequencies are added to each other with a positive sign. In YY
Her, where the tidal wave motion is retrograde, its frequency is added
to the fundamental frequency of the magnetic cycle with a negative sign.
We shall not make an attempt here to explain these findings.

Another difference between the two stars is that in BF Cyg the spin and
the orbital periods are modulating the system optical emission at
quiescence, as well as during outbursts. Not so for YY Her where the
orbital period manifests itself only at quiescence. The spin period is
modulating the emission during outbursts, although at some epochs it
also modulates  the quiescence emission through dark spots on the surface
of the giant. Table 1 summarizes the properties of BF Cyg and YY Her.

In paper I we proposed, following suggestions by other investigators
(e.g. Mikolajewska \& Kenyon 1992) that the main cause for the varying
optical luminosity of BF Cyg during outbursts is in variations in the
accretion rate onto the hot component of that system.
If this is also the case in YY Her, the strong
{\em P}3 modulation at the outbursts of this star implies that the
accretion
is modulated by the cool/donor component rotation frequency. However, if a
beamed outflow of matter from the rotating giant is the triggering agent of
an energy source in the vicinity of the hot component of the system, the
expected periodicity should be the synodic period of the giant. If indeed
this is the meaning of the {\em P}3= 551.4 d period, the sidereal rotation
period of the giant must be the {\em P}=2917 d period. On the other hand, the
secondary minima in the LC discussed in Section 3.8, and our interpretation
of them as manifestations of dark spots on the surface of the rotating
giant, would indicate that P3 is the sidereal rotation period of the star.
We shall not try to resolve this difficulty in the framework of this paper.

The fact that in BF Cyg the giant spin rate is slower while in YY Her it
is faster than the orbital angular velocity is of interest by itself. It
may be relevant for the study of synchronization
and circularization in close binary systems. It may also be of relevance
within the general area of the dynamical evolution of interacting binary
stars.

\section{Summary}

A 115 year light curve of the symbiotic star YY Her was analyzed. The
mean optical brightness of the system was found to be declining at a
constant rate of $\sim $ .01 magn in 1000 d. This suggests that YY
Her should be reclassified as a symbiotic nova.
The optical LC of the star is modulated by three independent periods,
{\em P}1=4650 d, {\em P}2=593.2 d and {\em P}3=551.4 d.
The first one is a mean time interval between successive
outbursts of the star, of which 7.5 events have been recorded. We
suggest that it is the period of a solar-type magnetic dynamo cycle
operating in the outer envelope of the giant star of this system. {\em P}2
is
the orbital revolution period. It modulates the luminosity of the star
at quiescence states of the system. {\em P}3 is the rotation period of the
giant star, modulating the emission during outbursts. At some quiescence
epochs it also modulates the quiescence emission, probably through dark
spots on the surface of the rotating giant. The difference between the
spin and the orbital frequencies excites a tidal wave in the atmosphere
of the giant. The LC of the system is also modulated at a frequency that
is a beat of the magnetic dynamo frequency and the tidal wave frequency.

YY Her is the third symbiotic star exhibiting outburst events that occur
at  nearly constant frequency, with a characteristic repetition
time of
a few thousands days. Similarly with the star BF Cyg, YY Her shows that
tides in the atmosphere of the magnetically active giant star affect
significantly the dynamo process operating in this star.

In contrast to BF Cyg, the deviation from locked synchronization in the
YY Her system is in the sense that the giant star is spinning with a
higher angular velocity than that of the orbital motion. In BF Cyg it is
the other way around.

\begin{tabular}{@{}lcrrrc@{}} \\
\multicolumn {6} {|c|}{\bf Table 1}{Comparison of the properties of BF Cyg
and YY Her}\\
\hline
  &   & BF Cyg &Ref.  & YY Her & Ref.   \\

\hline
\\
Giant Sp. Type    &     &  M5 III      & 1  & M4 III  & 1 \\
Luminosity L $\odot$    &  Lo &  5200  & 2    &   1100  & 2 \\
Binary period     &  {\em P}2 &  757.3      & 3    &   593.2 & 4\\
Giant Spin period&   {\em P}3 &  798.8      &3    &   551.4 & 4 \\
Tidal wave period&   {\em P}t & 14580       & 3   & -7825   &4 \\
Solar-type period &  {\em P}1 &  5375       & 3   & 4650    & 4 \\
Beat period       & {\em P}b  &  4436       & 3   & 2917    & 4 \\
\\
\hline
\end{tabular}
1) M\H{u}rset \& Schmid (1999)
2) M\H{u}rset et al. (1991)
3) Leibowitz \& Formiggini (2006)
4) This paper

\section*{Acknowledgments}

       We acknowledge with thanks the variable star observations from the
AAVSO
International Database contributed by observers worldwide and used in this
research.

This research is supported by ISF - Israel Science Foundation of the
Israeli Academy of Sciences.

\label{lastpage}

\end{document}